\documentstyle[epsf]{mn}
\newcommand{\mincir}{\raise
-2.truept\hbox{\rlap{\hbox{$\sim$}}\raise5.truept\hbox{$<$}\ }}
\newcommand{\magcir}{\raise
-2.truept\hbox{\rlap{\hbox{$\sim$}}\raise5.truept\hbox{$>$}\ }}
\newcommand{\minmag}{\raise
-2.truept\hbox{\rlap{\hbox{$<$}}\raise6.truept\hbox{$<$}\ }}
\newcommand{\be}{\begin{equation}}
\newcommand{\ee}{\end{equation}}
\newcommand{\ba}{\begin{eqnarray}}
\newcommand{\ea}{\end{esqnarray}}
\newcommand{\brr}{\begin{array}}
\newcommand{\err}{\end{array}}
\newcommand{\bc}{\begin{center}}
\newcommand{\ec}{\end{center}}

\title[Substructure - Alignment Connection]
{The Cluster Substructure - Alignment Connection}
\author[Plionis, Basilakos]{Manolis Plionis$^1$ \& Spyros Basilakos$^2$\\
$^1$Institute of Astronomy \& Astrophysics, National Observatory
of Athens, I.Metaxa \& B.Pavlou, Palaia Penteli, Athens 152 36, Greece \\
$^2$Astrophysics Group, Imperial College London, 
Blackett Laboratory, Prince Consort Road, London SW7 2BW, UK
}

\begin{document}

\maketitle

\begin{abstract}
Using a sample of 903 APM clusters we investigate whether their
dynamical status, as evidenced by the presence of significant substructures,
is related to the large-scale structure of the Universe.
We find that 
the cluster dynamical activity is strongly correlated with the
tendency of clusters to be aligned with their nearest neighbour and in
general with the nearby clusters that belong to the same
supercluster. Furthermore, dynamically active clusters are more 
clustered than the overall cluster population.
These are strong indications that clusters develop in a
hierarchical fashion by anisotropic merging along 
the large-scale filaments within which they are embedded.

\vspace{0.3cm}

\end{abstract}

\section{Introduction}
An interesting observable, that was thought initially to provide 
strong constraints on theories of galaxy formation, 
is the tendency of clusters to be aligned with their nearest 
neighbour as well as with other clusters that reside in the same 
supercluster (cf. Binggeli 1981; West 1989; Plionis 1994; 
Chambers, Melott \& Miller 2001). 
Analytical and numerical work have shown that such alignments, 
expected naturally to occur in "top-down" scenarios (cf. Zeldovich
1970), are also found in hierarchical clustering models of 
structure formation like the CDM (Bond 1986; West et al. 1991;
Splinter et al. 1997; Onuora \& Thomas 2000).
This fact could be explained as the result of an interesting 
property of Gaussian random fields that occurs for 
a wide range of initial conditions and which is the "cross-talk" between 
density fluctuations on different scales. 
Furthermore, there is strong evidence that the 
brightest galaxy (BCGs) in clusters is aligned
with the orientation of its parent cluster and even with the
orientation of the large-scale filamentary structure within which they
are embedded (cf. Struble 1990; West 1994, Fuller, West \& Bridges 1999). 

Within the framework of hierarchical clustering, the
anisotropic merger scenario of West (1994), in which clusters
form by accreting material along the filamentary structure within
which they are embedded, provides an interesting explanation of such
alignments as well as of the observed strong alignment of BCGs 
with their parent cluster orientation. In this framework, one should 
expect that dynamical young clusters, at there early stages of
formation in which they are not smooth and spherically symmetric, 
should show substructures that are aligned with the local
large-scale structures, an effect observed also in numerical
simulations for a variety of power-spectra 
(van Haarlem \& van de Weygaert 1993; Tormen 1997).
Indeed, supporting this view West, Jones \& 
Forman (1995) found, using 43 Einstein clusters that have a neighbour
within $\sim 10$ $h^{-1}$ Mpc, 
that cluster substructures do show a tendency to be aligned with the 
orientation of the major axis of their parent cluster and with 
the nearest-neighbouring cluster (see also Novikov et al 1999). 
In this work we investigate the relation between the strength of
cluster-cluster alignments and the large-scale environment in which the
clusters are embedded using the APM cluster catalogue (Dalton et al
1997), which is the largest one available.

\section{Methodology}
The APM cluster catalogue is based on the APM galaxy survey which
 covers an area of 4300 square degrees in the southern
sky containing about 2.5 million galaxies brighter
than a magnitude limit of $b_{J}=20.5$ (for details see Maddox et al. 1990).
Dalton et al (1997) applied an object cluster finding algorithm to the
APM galaxy data using a search radius of  $0.75 \; h^{-1}$ Mpc in order
to minimize projection effects, and so produced a list 
of 957 clusters with $z_{est} \mincir 0.13$. 
Out of these 309 ($\sim 32\%$) are ACO clusters, while 374 
($\sim 39\%$) have measured redshifts (179 of these are ACO clusters).
The APM clusters that are not in the ACO list are relatively poorer
systems than the Abell clusters, as we have verified comparing their 
APM richness's (see Dalton et al 1997 for definition of richness).

For the present analysis we use 903 of the above APM clusters,
since 54 clusters are found in the vicinity of plate-holes or 
crowded regions, a fact which affects severely their shape parameters.
The cluster distance is estimated from their redshift using 
$H_{\circ}=100 h$ km s$^{-1}$ Mpc$^{-1}$ and $q_{\circ}=0.5$.

\subsection{Cluster Shape Parameters and Alignment Measure}
A detailed analysis of the cluster shape determination procedure and of
the intrinsic APM cluster shapes can be found in Basilakos, Plionis \&
Maddox (2000). Here we only sketch the basic procedure which is based
on the familiar moments of inertia method with 
$I_{11}=\sum\ w_{i}(r_{i}^{2}-x_{i}^{2})$,
$I_{22}=\sum\ w_{i}(r_{i}^{2}-y_{i}^{2})$,
$I_{12}=I_{21}=-\sum\ w_{i}x_{i}y_{i}$,
where $x_i$ 
and $y_i$ are the Cartesian coordinates of the galaxies that their
projected separation is such that they are judged as belonging to the
cluster (details in Basilakos et al 2000)
and $w_i$ is their weight. We, then diagonalize 
the inertia tensor solving the basic equation:
\begin{equation}\label{eq:diag}
det(I_{ij}-\lambda^{2} \; M_{2})=0 ,
\end{equation}
where $M_{2}$ is the $2 \times 2$ unit matrix. The cluster ellipticity
is given by
$\epsilon=1-\frac{\lambda_2}{\lambda_1}$, where $\lambda_i$ are the
 positive eigenvalues with $(\lambda_1>\lambda_2)$.
This method can be applied to the data using either the discrete or smoothed
distribution of galaxies. The determination of the 
cluster orientation is consistent among the two methods 
but this is not true also for the cluster
ellipticity (for details see Basilakos et al. 2000)
\begin{figure}\label{fig:alin}
\epsfxsize=8.5cm \epsffile{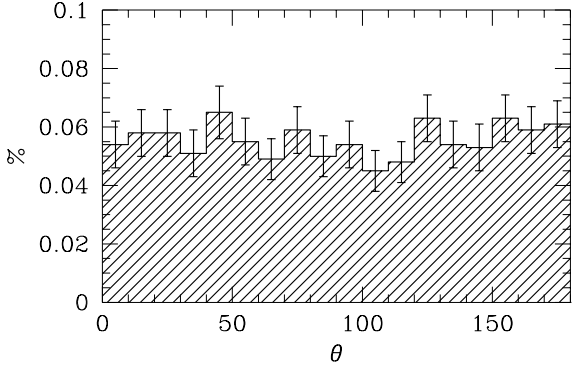}
\caption{Distribution of APM cluster position angles.}
\end{figure}

In Figure 1 we present the derived cluster position angle, $\theta_i$, 
distribution. We see no significant systematic orientation effects, a
fact that we quantify estimating the Fourier transform of the galaxy
position angles;
$C_{n} = \sqrt{\frac{2}{N}} \sum \cos 2n\theta_i$
and
$S_{n} = \sqrt{\frac{2}{N}} \sum \sin 2n\theta_i$.
If the galaxy position angles are uniformly distributed between $0^{\circ}$ 
and $180^{\circ}$, then both $C_{n}$ and $S_{n}$ have zero mean and unit 
standard deviation. Therefore large values ($>2.5-3$) indicate significant 
deviation from isotropy. We find that $C_{1,2}$ and $S_{1,2}$ have values
$\mincir 1.2$, which indicates that there is no systematic orientation
bias. 

In order to investigate the alignment between cluster orientations,
we define the relative position angle
between the major axis orientation of a cluster and the direction 
to a neighbouring one 
by $\phi_{i,j}\equiv |\theta_i - \vartheta_j|$
(where $\vartheta$ is the position angle of the cluster pair separation
vector).
In an isotropic distribution we will have
$\langle \phi_{i,j} \rangle \simeq 45^{\circ}$. 
A significant deviation from this would be an indication of an 
anisotropic distribution which can be quantified by 
(Struble \& Peebles 1995):
\begin{equation}\label{eq:alin}
\delta=\sum_{i=1}^{N}\frac{\phi_{i,j}}{N}-45
\end{equation}
In an isotropic distribution we have $\langle \delta \rangle \simeq 0$, while
the standard deviation is given by $\sigma=90/\sqrt{12 N}$. 
A significantly negative value of $\delta$ would indicate alignment and 
a positive misalignment.
To avoid problems related 
to ill defined position angles we will use in our analysis only clusters
that have ellipticities $>0.05$ (ie., $N=888$)\footnote{Our results
remain unaltered even for higher values of the ellipticity cutoff.}.
It should be noted that systematic biases and projection 
effects will tend to mask any true alignment signal. For example,
the projection of foreground galaxies along the line of sight of a cluster 
as well as the projection on the plane of the sky of member galaxies,
always work in the direction of smearing alignments.

\subsection{Substructure Measure \& Significance}
Major obstacles in attempting to determine the dynamical state of a
cluster is (1) the ambiguity in identifying cluster
substructure in 2D or even 3D cluster data and 
(2) the uncertainty of post-merging relaxation timescales.
Evrard et al. (1993) and Mohr et al. (1995)
have suggested as an efficient
indicator of cluster substructure the shift of the center-of-mass 
position as a function of density threshold above which 
it is estimated. The {\em centroid-shift} ($sc$) is defined
as the distance between the cluster center-of-mass, 
$(x_{\rm o}, y_{\rm o})$, which may change at different density
thresholds and the highest cluster density-peak,
$(x_{\rm p}, y_{\rm p})$, ie., 
$sc = \sqrt{(x_{\rm o}-x_{\rm p})^{2}\,+\,(x_{\rm o}-x_{\rm p})^{2}}$.

Kolokotronis et al. (2001), using in a 
complementary fashion optical and X--ray data (see also
Rizza et al. 1998), since in the X-ray band
projection effects are minimal, calibrated various substructure measures
using APM data and pointed ROSAT observations of 22 Abell clusters
and found that in most cases using X--ray or optical data one can
identify substructure unambiguously. Only in $\sim 20\%$ of the
clusters that they studied did they find projection effects in
the optical that altered the X-ray definition of substructure. 
An important conclusion of Kolokotronis et al. (2001)
was that a large and significant value of $sc$ is a clear indication 
of substructure in APM optical cluster data.

The significance of such centroid variations to
the presence of background contamination and random density
fluctuations are quantified using 
Monte Carlo cluster simulations in which, by construction, there is no
substructure. For each APM cluster, 1000 simulated
clusters are produced having the observed ellipticity, 
the observed number of galaxies, following a King's profile, as well as 
a random distribution of expected background galaxies, determined by the
distance of the cluster and the APM selection function (note that the
number of ``galaxies'' that we allow to follow 
the King's profile is the observed number minus the expected random
background).
The King-like profile is:
\begin{equation}\label{eq:sb}
\Sigma(r) \propto \left[1\,+\,\left(\frac{r}{r_{\rm c}} \right)^{2}
\right]^{-\alpha} \;,
\end{equation}

\noindent 
where $r_{\rm c}$ is the core radius. We use the weighted, by the
sample size, mean of 
most recent $r_c$ and $\alpha$ determinations 
(cf. Girardi et al. 1998),
i.e., $r_{\rm c} \simeq 0.085 \;h^{-1}$ Mpc and $\alpha \simeq 0.7$. 
We do test the robustness of our results for a plausible range of these
parameters (details can be found in Kolokotronis et al. 2001).
 
Naturally, we expect the simulated clusters to generate
small $sc$'s and in any case insignificant shifts.
Therefore, from each set of Monte-Carlo cluster simulations 
we derive $\langle sc \rangle_{\rm sim}$
as a function of the same density thresholds
as in the real cluster case. Then, within a search radius of
$0.75 \;h^{-1}$ Mpc from the simulated highest cluster peak,
we calculate the quantity:
\begin{equation}\label{eq:sig}
\sigma =\frac{\langle sc \rangle_{\rm o} - \langle sc 
\rangle_{\rm sim}}{\sigma_{\rm sim}}\;,
\end{equation}
which is a measure of the significance of real centroid shifts
as compared to the simulated, substructure-free clusters. Note that 
$\langle sc \rangle_{\rm o}$
is the average, over three density thresholds, centroid shift
for the real APM cluster. 

A further possible substructure identification procedure is based
on a friend-of-friends 
algorithm, applied on 3 overdensity thresholds of each cluster
(for details see Kolokotronis et al. 2001). Three
categories are identified, based on the subgroup multiplicity and
size: {\em (a)} No substructure (unimodal), {\em (b)} 
Weak substructure (multipole groups but with total group mass 
$\le 25\%$ of main), 
{\em (c)} Strong substructure (multipole groups but with mass $>
25\%$ of main). 

\section{Results \& Discussion}
\subsection{Cluster Substructure and Alignments}
Applying the {\em centroid shift} substructure identification procedure
to the 903 APM clusters we find that
about 30\% of clusters have significant ($> 3 \sigma$)
substructure. Note that
defining as having significant substructure those clusters with $\sigma
>2.5$ or 2 increases the fraction to $\sim$ 40\% and 50\%
respectively. Furthermore, changing the structural 
parameters of the Monte-Carlo
clusters changes the actual $\sigma$-values, although their relative
significance rank-order remains unaltered. 
Alternatively if we apply the {\em subgroup} categorization procedure
we find that $\sim$53\% of the APM clusters show strong indications of
substructure. 

We have tested whether there is any 
systematic redshift dependent
effect of the cluster substructure categorization and found none (using
either measured or estimated redshifts).
We conclude that irrespectively of the method $\sim 30\%
- 50\%$ of the APM clusters show indications of significant
substructure (in accordance with Kolokotronis et al 2001).

We now test whether the well known nearest-neighbour alignment
effect, present in the Abell clusters (cf. Bingelli 1982; Plionis 1994), 
is evident also in the poorer APM clusters.
In table 1 we present, for the whole APM cluster sample, our alignment results 
as a function of maximum intercluster separation, $D_{max}$, for both
the nearest-neighbours ($\delta_{nn}$) and for all neighbouring pairs
($\delta_{an}$), within $D_{max}$.

\begin{table}
\tabcolsep 3pt
\begin{tabular}{ccccccc} \hline 
$D_{max}$ & $\langle \delta_{nn} \rangle$ & $P(>\chi^2)$ & $N_{pairs}$
& $\langle \delta_{an} \rangle$ & $P(>\chi^2)$ & $N_{pairs}$\\ \hline
5   & -7.7$\pm 2.7$ & 0.005 & 90 & -7.5 $\pm 2.5$ & 0.002 &110 \\
10  & -4.6$\pm 1.6$ & 0.005 & 270 & -4.1$\pm 1.3$ & 0.000 & 374 \\
15  & -3.8$\pm 1.2$ & 0.001 & 444 & -2.6$\pm 0.9$ & 0.002 & 848 \\
20  & -2.9$\pm 1.0$ & 0.004 & 616 & -1.1$\pm 0.7$ & 0.023 & 1602 \\
30  & -1.7$\pm 0.9$ & 0.069 & 805 &  -0.9$\pm 0.4$ & 0.112 & 4012 \\ \hline
\end{tabular}
\caption{Nearest neighbour ({\em nn}) and all-neighbour ({\em an}) 
alignment signal as a function of maximum pair separation, $D_{max}$. 
The $\chi^2$ probabilities 
are derived by comparing the binned $\phi$-distribution (using 3 bins of
30$^{\circ}$ width) with the Poisson expectation values.}
\end{table}
It is evident that there is significant 
indication of cluster alignments, with the alignment signal dropping 
in amplitude and significance, as a function
of increasing $D_{max}$. In order to test whether this
result is dominated by the ACO cluster pairs, and thus whether it is a
manifestation of the already known Abell cluster alignment effect,
we have excluded such pairs (113/888 for the $nn$-case) to find consistent 
but more significant alignment results. For 
example for the $D_{max}=15$ $h^{-1}$ Mpc case, we obtain
$\langle \delta_{nn} \rangle \simeq -4.3 \pm 1.3$ with
$P(>\chi^2)=0.0004$ and $\langle \delta_{an} \rangle \simeq -2.7 \pm 0.9$ with
$P(>\chi^2)=0.001$.

Note that a large number of APM clusters have estimated redshifts and thus
the cluster distance uncertainties will tend to hide true
alignments. Therefore, the measured alignment signal should be considered
rather as a lower limit to the true one.

\subsection{The Substructure-Alignment Connection}
We have correlated the alignment signal with the substructure
significance indication in order to see whether there is any relation
between the large-scale environment, in which the cluster distribution
is embedded, and the internal cluster dynamics. 
\begin{figure}\label{fig:alinsig}
\epsfxsize=8.4cm \epsffile{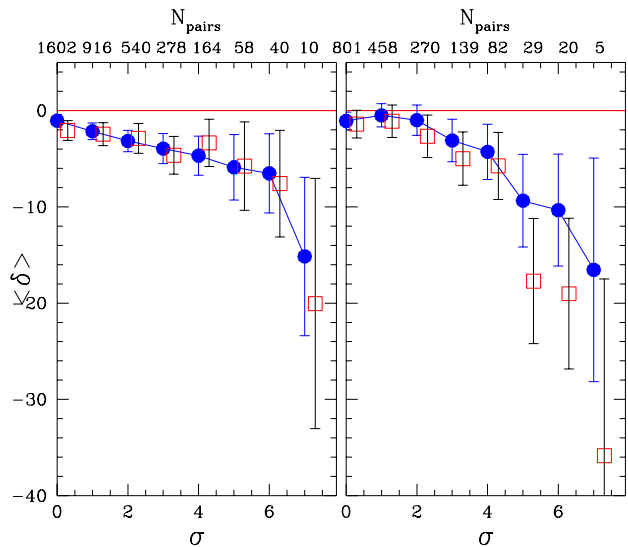}
\caption{Alignment signal of all cluster pairs with separation
$D_{cc}\le 20$ $h^{-1}$ Mpc, as a function of substructure 
significance, $\sigma$. The left panel presents alignments between
cluster major axis orientation and the direction to a neighbour while
the right panel represents the alignments between the two cluster major
axes orientation. Furthermore, the filled symbols represent the signal
based only on the {\em centroid-shift} substructure categorization
while the open symbols represent the signal from clusters that are also
categorized as having {\em strong} substructure 
by the {\em subgroup} categorization procedure.}
\end{figure}

In figure 2 we present the alignment signal, 
$\langle \delta \rangle$, between all cluster pairs
with separations $< 20$ $h^{-1}$ Mpc that have substructure
significance above the indicated $\sigma$ value.
Evidently, there is a strong correlation
between the strength of the alignment signal and the substructure
significance level (the two panels are based on slightly different
definition of the alignment signal - see caption for details).

In order to assess the statistical significance of
obtaining such a $\delta - \sigma$ relation, we
perform a large number (10000) of Monte-Carlo simulations in which we
reshuffle the measured position angles, assigning them at random to
the clusters. Then we derive the
corresponding $\delta - \sigma$ relation and count in how many
such simulations do we get values of $\delta$ which are as
negative, or more, than the corresponding ones of figure 2. We find
that the corresponding probability is $<10^{-5}$, while for the less
restrictive case, of having negative $\delta$-values (of any amplitude) 
for all $\sigma$, the probability is again quite small ($\sim 7.5 \times
10^{-3}$). 

Note that from the analysis of Kolokotronis et al
(2001) it is expected that our procedure will misidentify  
the dynamical state of $\sim 20\%$ of the APM clusters. However, such 
misidentification will act as a noise factor
and will tend to smear any true alignment-substructure correlation,
since there is no physical reason why random projection effects, within
0.75 $h^{-1}$ Mpc of the cluster core, should be correlated with the
direction of neighbours within distances up to a few tens of
Mpc's (such a correlation could be expected at some level only for
nearest-neighbours in angular space but we have verified that choosing 
such pairs we obtain an insignificant alignment signal).
Therefore our result, based on the largest cluster sample
available, supports the hierarchical clustering scenario and in
particular the formation of cluster by anisotropic merging along the
filamentary structure within which they are embedded 
(cf. West 1994; West, Jones \& Forman 1995). 
\begin{figure}\label{fig:xi}
\epsfxsize=9.cm \epsffile{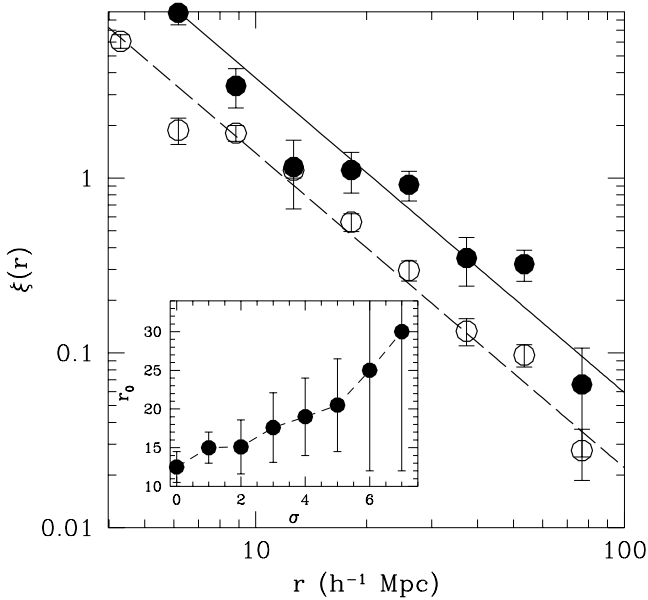}
\caption{Two-point correlation function of all APM clusters
(open symbols) and of $\sigma>4$ clusters (filled symbols). The lines
represent the best $(r/r_{\circ})^{-1.8}$ fit with $r_{\circ}\simeq12$ and
$\simeq 20$ $h^{-1}$ Mpc respectively. 
Insert: The cluster correlation length
as a function of substructure significance.}
\end{figure}

\subsection{Local density - Substructure Correlation}
If the above view is correct then one would expect that clusters with
significant substructure should be residing preferentially in
high-density environments (superclusters), and this would then
have an imprint in their spatial two-point correlation function. 

In figure 3 we present the spatial 2-point
correlation function of all APM clusters (open symbols) and of those with
substructure significance $\sigma \ge 4$ (filled symbols). It is clear that
the latter are significantly more clustered. This can be seen also in
the insert of figure 3 were we plot the correlation length, $r_{0}$, as a
function of $\sigma$, which is clearly an increasing function of
cluster substructure significance level. Note that in order to 
take into account
the possible systematic distance dependent effects in the different 
cluster subsamples we generate random catalogues, used to normalize 
the number of cluster-cluster pairs in the $\xi(r)$ estimate, 
using the individual distance distribution of each
subsample and not the overall APM cluster selection function. Had we
used the latter, the increase of $r_{\circ}$ with $\sigma$ would have
been more severe.

Furthermore, we have tested whether this effect
could be due to the well-known richness dependence of the correlation strength,
and found a weak, if any, such richness trend. 
In order to further investigate this issue we have correlated
the APM richness (see Dalton et al 1997) with $\sigma$ and found a very
weak but significant correlation (the Pearson's coefficient is only 
0.26 but with a significance $>$99.9\%). Therefore, a component
contributing to the increase of the 
$r_{\circ}(\sigma)$ function could possibly be the cluster
richness effect (through the above richness-$\sigma$ weak correlation).
However, the increase of $r_{\circ}(\sigma)$ is quite dramatic and
cannot be attributed only to this weak richness dependence.
This is corroborated also from the fact that there is a larger
fraction of clusters with significant substructure 
(high $\sigma$ values) residing in high-density regions
(superclusters), as shown in figure 4 where 
we plot the fraction of clusters with
$\sigma>3$ that belong to superclusters identified by the indicated
percolation radius. Evidently the fraction increases inversely
with percolation radius. These results are similar to those of
Schombert \& West (1990) based on Abell clusters, in which they found
that flattened clusters, flatness being an indication of dynamical youth,
are more frequent in high density environments
(see however Herrera \& Sanroma 1997 for a different view).
\begin{figure}
\epsfxsize=10.5cm \epsffile{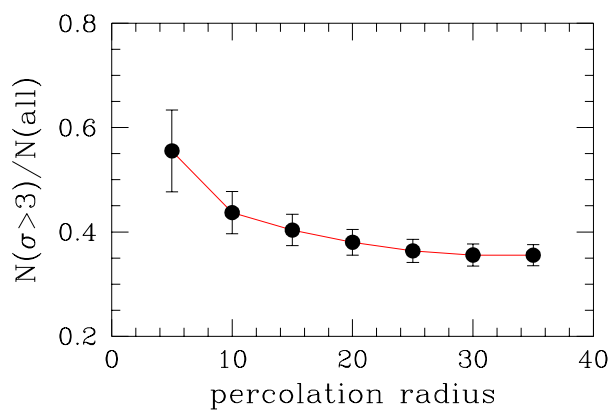}
\caption{The fraction of clusters with significant substructure
($\sigma > 3)$ that
reside in superclusters, defined by the indicated percolation radius.}
\end{figure}

The conclusion
of this correlation function analysis is that indeed the clusters
showing evidence of dynamical activity reside in high-density
environments, as anticipated from the alignment analysis. 
It is interesting that such environmental dependence has also been found 
in a similar study of the BCS and REFLEX clusters 
(Sch\"{u}ecker et al. 2001) and for
the cooling flow clusters with high mass accretion rates (Loken, Melott
\& Miller 1999). 

\section{Conclusions}
We have presented evidence, based on the largest available cluster
sample, the APM, that there is a strong link between the 
dynamical state of clusters and their large-scale environment.
Clusters showing evidence of
dynamical activity are significantly more aligned with their nearest
neighbours and they are also much more spatially clustered.
This supports the hierarchical clustering models in which clusters
develop by accreting matter along the large-scale filamentary
structures within which they are embedded.

\section*{Acknowledgements}
M.Plionis acknowledges the hospitality of the Astrophysics Group
of Imperial College.
This work was partially supported by the EC Network programme `POE'
(grant number HPRN-CT-2000-00138). We thank the referee, M.West, for
very helpful suggestions that improved our work.

\end{document}